\documentclass[aps,pra,notitlepage,superscriptaddress,twocolumn,raggedbottom,longbibliography,10pt]{revtex4-1}
\usepackage[colorlinks=true, allcolors=blue]{hyperref}		
\usepackage{amsmath,bm,amssymb,esint}			
\usepackage{lmodern}							
\usepackage{enumitem}							
\usepackage{graphicx}							
\usepackage{lineno}								
\usepackage{xcolor}								
\usepackage[UKenglish]{babel}					
	\usepackage{microtype}							

\newcommand{\nn}{\nonumber}
\newcommand{\bx}{\ensuremath{\mathbf{x}}}
\newcommand{\bn}{\ensuremath{\mathbf{n}}}
\newcommand{\avg}[1]{\left\langle{#1}\right\rangle}

\begin{document}
\title{Consensus and diversity in multi-state noisy voter models}

\author{Francisco Herrer\'{i}as-Azcu\'{e}}
	\email{francisco.herreriasazcue@manchester.ac.uk}
	\affiliation{Theoretical Physics, School of Physics and Astronomy, The University of Manchester, Manchester M13 9PL, United Kingdom}
\author{Tobias Galla}
	\email{tobias.galla@manchester.ac.uk}
	\affiliation{Theoretical Physics, School of Physics and Astronomy, The University of Manchester, Manchester M13 9PL, United Kingdom}

		\begin{abstract}

We study a variant of the voter model with multiple opinions; individuals can imitate each other and also change their opinion randomly in mutation events.
We focus on the case of a population with all-to-all interaction.
A noise-driven transition between regimes with multi-modal and unimodal stationary distributions is observed. In the former, the population is mostly in consensus states; in the latter opinions are mixed. 
We derive an effective death-birth process, describing the dynamics from the perspective of one of the opinions, and use it to analytically compute marginals of the stationary distribution. 
These calculations are exact for models with homogeneous imitation and mutation rates, and an approximation if rates are heterogeneous.
Our approach can be used to characterize the noise-driven transition and to obtain mean switching times between consensus states.
		\end{abstract}


\maketitle
 

\section{Introduction}\label{sec:Intro}
The Voter Model (VM) was initially introduced as a model for spatial conflict \cite{Clifford1973, Liggett2012interacting}. In its most basic variant, the model describes a population of voters with pairwise interactions. Each individual can be in one of two states. In an interaction, one individual copies the state of another. If the population is finite, this process continues until all individuals have reached the same state; no further dynamics are then possible.

This model has been studied in the context of different applications. For example, variants of the VM have been used to describe autocatalytic reactions \cite{Frachebourg1996}, herding in financial markets \cite{Kirman1993}, opinion dynamics \cite{Castellano2009}, and the evolution of language \cite{Castello2006,Castellano2009}. In the context of biological evolution, the VM is closely related to the Moran process with neutral selection \cite{Granovsky1995}. In each of these applications the states an individual can take represent different properties. They can be opinions on a given issue, trading behaviors in a model of a financial market, or different species in a biological context. 

Although the models used in these applications share common characteristics, different research communities have focused on different aspects of the VM and its applications. For the purpose of opinion dynamics, for example, one may be interested in whether or not a population of agents reaches consensus and, if they do, how long this takes. In economics one may ask how copying of trading behaviour leads to herding, and in the context of linguistics one might be interested in understanding how features of languages spread and organise in space \cite{Kauhanen2018}. Applications of the VM include the use to model actual elections \cite{FernandezGracia2014}.

The VM is also of interest from a point of view of statistical mechanics. The model is similar to spin models in traditional statistical physics, except that it is defined dynamically and not through an energy function. As such, the VM has become one of the most studied models of non-equilibrium statistical physics. The two-state VM in two dimensions has been shown to exhibit logarithmic coarsening \cite{Dornic2001}. This is in contrast for example with Glauber dynamics for the Ising model, in which coarsening is driven by surface tension and growth laws are algebraic. Multi-state voter models, on the other hand, can show algebraic coarsening \cite{DallAsta2008}. 
The traditional VM presents an interesting class of dynamics, with two absorbing states and $\mathbb{Z}_2$-symmetry. It has been shown to define its own universality class \cite{Dornic2001,AlHammal2005}, and field theories have been devised to study its critical point \cite{Dornic2005}. 
A further aspect attracting attention in the physics community has been the coupled dynamics of the voter model and interaction networks between agents. Such co-evolutionary processes have, been shown to lead to fragmentation transitions \cite{Holme2006,Vazquez2008a}, enriching the number and nature of the absorbing states of the VM.

One interesting variant in the class of voter dynamics is the so-called `noisy voter model' \cite{Granovsky1995}. The terminology might at first sound surprising -- the process of opinion changes through interaction in the original VM is already stochastic. However, it allows for absorbing states, in which no further dynamics can occur.

What is meant by the term `noisy' in this context is that, in addition to the interaction with other individuals, random changes of opinion can occur; they do not require interaction with another agent, and are sometimes also described as `mutations'. These random changes drive the system away from consensus; there are no absorbing states. 

As a consequence, two effects compete in the noisy voter model: a drive towards consensus through interaction, and spontaneous opinion changes promoting coexistence. This leads to a transition between a regime in which the system is mostly ordered (i.e., all agents are of the same opinion), and another regime in which both opinions are represented in the system. This phenomenon is known as `noise-induced bistability'  and has been investigated in the context of autocatalytic reactions, surface-reaction models, decision making of insects, and biochemical reactions \cite{Fichthorn1989, Togashi2001, Ohkubo2008, Biancalani2012, Biancalani2014, Houchmandzadeh2015, Saito2016}. Recently, transitions of this type have also been studied in noisy voter models on complex networks \cite{Carro2016,Peralta2018a,Peralta2018}.

Most of the existing work on models of this type focuses on the case in which individuals can take two different states; the transition is then between a phase in which the stationary distribution of individuals is unimodal and another in which it is bimodal; hence the term `noise-induced bistability'. 

The aim of the present work is to generalise the model to the case of multiple opinion states. In the absence of mutation there are then multiple absorbing states. As we demonstrate, the inclusion of spontaneous state changes leads to noise-induced multi-stability. We investigate this numerically and analytically. To do this, we compute marginals of the stationary distribution of the model, as well as switching times between the different consensus states. Our analysis focuses on the case of a `mean-field' geometry, that is, a population in which all pairs of individuals can interact at all times.

The remainder of the paper is organized as follows. In Sec.~\ref{sec:model} we define the model and introduce the general notation. In Sec.~\ref{sec:reduction} we derive an effective master equation for the dynamics of a single species and use it to describe marginals of the stationary distribution of the multi-species model; closed-form analytical expressions are obtained. This approach is exact if all species have the same imitation and mutation rates, and an approximation otherwise. We also estimate switching times between different consensus states. Sec.~\ref{sec:hom} focuses on the model with homogeneous imitation and mutation rates. We obtain analytical predictions for its phase diagram and test these predictions against simulations. In Sec.~\ref{sec:het} we carry out a similar analysis for a multi-state model with heterogeneous rates. Sec.~\ref{sec:concl} contains a discussion of our results and an outlook on future work.

\section{Definition of the model and noise-induced multi-stability}\label{sec:model}

\subsection{Model definition and notation}
We consider a population of $N$ individuals; each can be of one of $m$ types ($m\in\mathbb{N}$). In the context of social dynamics these would represent different opinions; in evolutionary biology they may stand for different species. Our aim is not to study a specific application, but the general structure that emerges from multi-state noisy voter models; we will therefore use the terms opinions, species and types interchangeably. 

The population is unstructured, that is, any individual can interact with any other member of the population at all times. We write $n_i$ for the number of individuals of species $i$, where ${i\in\{1,\dots,m\}}$; these variables vary as the population evolves. At each time the state of the system is fully specified by the vector ${\bn=(n_1,\dots,n_m)}$. The size of the population is constant, i.e., we have $\sum_{i=1}^m n_i=N$ at all times; the state space is a simplex in $m$-dimensions.

The dynamics of the model are defined by the following imitation and mutation reactions:
\begin{align}\label{eq:reactions}
	X_{i}+X_{j} & \xrightarrow{r_{ji}} 	 		2X_{i} \nn\\
	X_{j} 		& \xrightarrow{\epsilon_{ji}}	X_{i}.
\end{align}
The notation $X_i$ represents an individual of species $i$, and the first reaction describes the process in which an individual of opinion $j$ imitates an individual of opinion~$i$. We write $r_{ji}$ for the corresponding imitation rate; the exact interpretation of these coefficients will be made more precise below in Eq.~(\ref{eq:rates}). In the context of evolutionary dynamics reactions of this type represent a death-birth event: an individual of type $j$ dies, and is replaced by the offspring of an individual of species $i$. This offspring is also of type $i$. To keep the model general, we allow this rate to depend on both species involved, $j$ and $i$. 

The second reaction in Eq. (\ref{eq:reactions}) describes spontaneous changes of opinion, i.e., an individual of type $j$ turns into an individual of type $i$. The rate with which this process occurs is $\varepsilon_{ji}$, where we again include possible dependence on $j$ and $i$.  In the context of population dynamics it may be more realistic to incorporate mutation in the reproduction events; that is, one could consider models in which the offspring of an individual does not necessarily belong to the same species as the parent. The effect of both ways of introducing random state changes is similar though: they allow departure from states in which the entire population is made up of individuals of one single species. We will refer to instances of the second process in Eq.~(\ref{eq:reactions}) as `mutation events'.

The dynamics are defined in continuous time, and can be described by the master equation
\begin{equation}
	\partial_{t}P\left(\bn,t\right)=\sum_{i=1}^{m}\sum_{j=1}^{m}
		\left({\cal E}_{i}^{-1}{\cal E}_{j}-1\right)\left[T_{j\to i}(\bn)P\left(\bn,t\right)\right],
	\label{eq:master}
\end{equation}
where $P(\bn,t)$ is the probability of finding the population in state $\bn$ at time $t$. We have written ${\cal E}_i$ for the creation operator for individuals of species $i$; it is defined by its action ${\cal E}_i f(n_1,\dots,n_i,\dots,n_m)=f(n_1,\dots, n_i+1,\dots,n_m)$ on functions $f(\bn)$ of the state of the population. The quantity $T_{j\to i}(\bn)$ is the rate with which individuals of type $j$ are converted into individuals of type $i$ if the system is currently in state $\bn$; these rates are given by 
\begin{equation}
	T_{j\rightarrow i}(\bn)=r_{ji}\frac{n_{i}n_{j}}{N}+\epsilon_{ji}n_{j}.	\label{eq:rates}
\end{equation}

\begin{figure*}[t!]
	\center
	\includegraphics[width=0.9\textwidth]{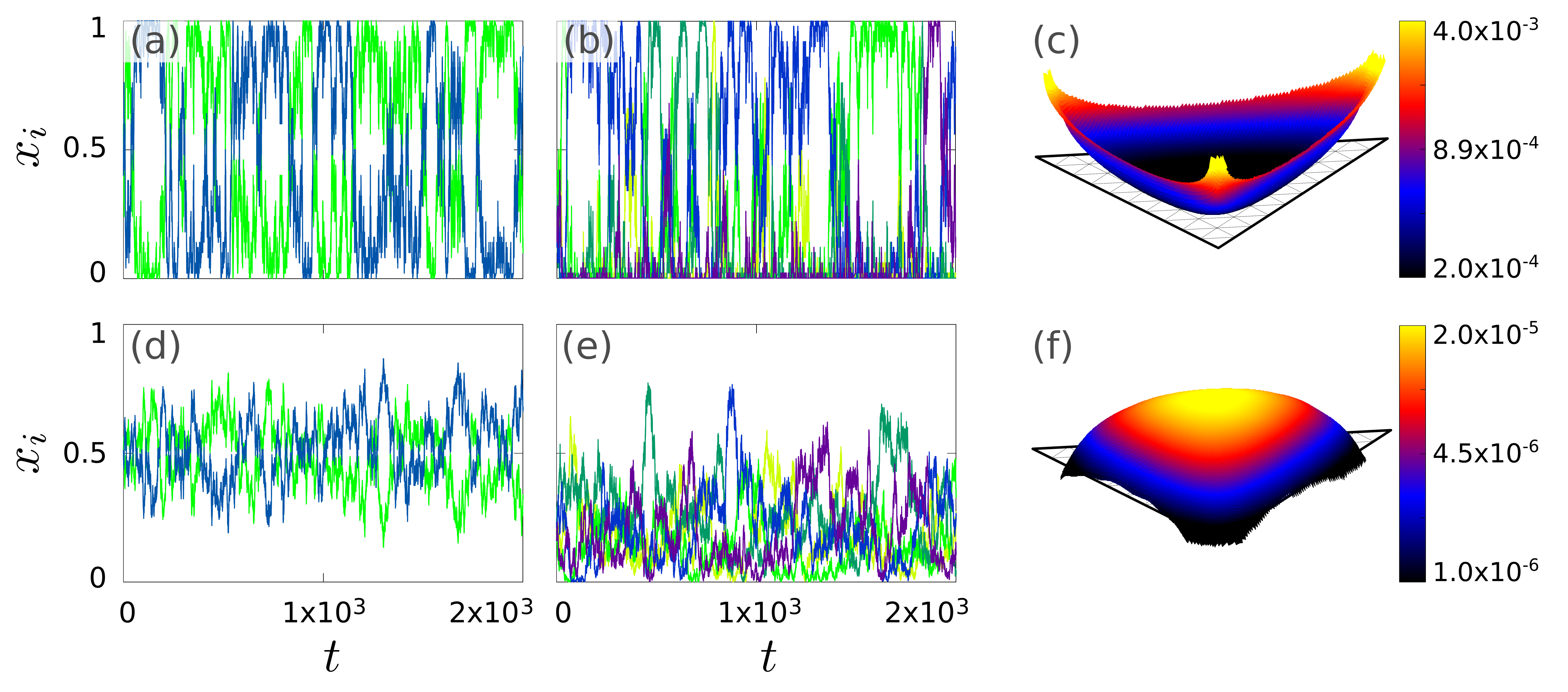}
	\caption{\textbf{Sample trajectories and stationary distribution.}
	Panels (a), (b), (d) and (e) show single realizations of the model dynamics; the distributions in panels (c) and (f) are from an average over many realizations. Panels (a) and (d) are for $m=2$; (b) and (e) for $m=5$; (c) and (f) for $m=3$. The upper panels (a)--(c) are for a population size of $N=50$; the system is frequently in states of full consensus. In the lower panels (d)--(f) $N=500$, and diversity of opinions is observed; states of consensus are rarely visited. The imitation and mutation rates are uniform across species; we use $r=1$ and $\varepsilon=10^{-2}/(m-1)$.} 
		\label{fig:JustVote}
\end{figure*}

We note that this is the overall rate for such events in the population, and not a per capita rate. As it is common practice, the scaling with the population size, $N$, is such that $T_{j\to i}={\cal O}(N)$. This ensures that ${\cal O}(N)$ events occur per unit time; in other words, time is measured in generations, rather than individual events. 

Among the states of the population there are states of complete consensus, in the language of opinion dynamics. This is the case when $n_i=N$ for one opinion $i$, and $n_j=0$ for all $j\neq i$. In the context of evolutionary dynamics these are monomorphic states; one species has taken over the entire population. These consensus states are not absorbing when mutation rates $\varepsilon_{ji}$ are non-zero. As we will discuss next, it is the interplay between imitation, mutation and the associated intrinsic noise that generates the interesting behaviour of the model.

\subsection{Noise-induced multistability}
Simulations of the model can be carried out efficiently using the standard Gillespie algorithm \cite{Gillespie1976, Gillespie1977}. Different types of outcomes are illustrated in Fig.~\ref{fig:JustVote}. We show the time evolution of the fraction of individuals in each species, $x_i=n_i/N$, in individual simulations of models with $m=2$ and $m=5$ species. In panels  (a)--(c) the population is of size $N=50$, and the population is seen to visit states of complete consensus relatively frequently. These are not permanent; instead, the population switches between different consensus states. In panels (d)--(f) the population size is $N=500$. Consensus may occasionally be reached, but other than that the system is mostly found in states of diversity, where different opinions are represented at fluctuating frequencies. Panels (c) and (f)  show the multi-modal and unimodal stationary distributions for a model with $m=3$.

\section{Reduction to effective single-opinion dynamics}\label{sec:reduction}
We proceed to characterise the multi-state noisy VM analytically. To do this, we first derive an effective death-birth process for the individuals of a particular species. We then study the shape of the stationary distribution resulting from these effective dynamics. We also use the single-species dynamics to obtain information about typical switching times between consensus states.

\subsection{Effective single-species master equation}
We focus on a particular species $i$ and the dynamics of the number of individuals $n_i$ of this species. Events in the population dynamics convert individuals of one species into individuals of another. Many of these events will not involve species $i$, and are therefore not relevant for the effective dynamics of species $i$. In the following, we only consider events which either increase or decrease $n_i$. 

The overall rate with which individuals are converted to species $i$ is 
	\begin{equation}\label{eq:tplus}
	T_{i}^+(\bn)\equiv \sum_{j\neq i} T_{j\to i}(\bn),
	\end{equation}
where $T_{j\to i}(\bn)$ is the conversion rate from $j$ to $i$, as defined in Eq.~(\ref{eq:rates}). Similarly, the rate with which individuals of species $i$ are converted into any other type is
	\begin{equation}\label{eq:tminus}
	T_{i}^-(\bn)\equiv \sum_{j\neq i} T_{i\to j}(\bn).
	\end{equation}
We note that these rates are in general not specified by $n_i$ alone; instead they depend on the entire state vector $\bn$. The rates in Eqs.~(\ref{eq:tplus}) and (\ref{eq:tminus}) by themselves, therefore, do not describe a well-defined stochastic process for species~$i$.

Inserting the definitions from Eq.~(\ref{eq:rates}) into the expressions in Eqs.~(\ref{eq:tplus}) and (\ref{eq:tminus}), one finds 
	\begin{align}\label{eq:tpm}
	T_{i}^+(\bn)&=\sum_{j\neq i}\left( r_{ji}\frac{n_in_j}{N}+\varepsilon_{ji} n_j\right),\nn \\
	T_{i}^-(\bn)&=\sum_{j\neq i} \left(r_{ij}\frac{n_in_j}{N}+\varepsilon_{ij} n_i\right).
	\end{align}
	
Our analysis in following sections is based on the assumption that we can formulate a closed death-birth process for species $i$, with birth and death rates $T_i^\pm(n_i)$, dependent only on $n_i$. As we will describe below, an exact description of this form can be obtained for the case of homogeneous rates across opinions (see Sec.~\ref{sec:hom}). If the imitation or mutation rates are heterogeneous this approach constitutes an approximation (see Sec.~\ref{sec:het}).

\subsection{Stationary distribution for individual opinions}\label{sec:StatDist}
At long times the $m$-species distribution $P(\bn,t)$ becomes time-independent; we write $P_{\rm st}(\bn)$ for the stationary distribution. In principle, this stationary distribution can be obtained from the linear set of equations for $P_{\rm st}$, obtained by setting the right-hand side of the master equation (\ref{eq:master}) to zero. In practice, it is very difficult to evaluate this analytically. However, as we will describe next, we can obtain the marginal distributions $P_i(n_i)=\sum_{\bn_{-i}} P_{\rm st}(\bn)$ in some cases; the notation $\sum_{\bn_{-i}}\dots$ in this expression indicates a sum over all variables $n_1,\dots,n_m$, except $n_i$. 

We assume that we can formulate a closed death-birth process for species $i$, with birth and death rates $T_i^\pm(n_i)$. The stationary distribution of this process, is then the marginal $P_i$, and it is given by
\begin{equation}\label{eq:statp}
P_i(n_i)=\frac{\prod_{k=1}^{n_i} \frac{T_i^+(k-1)}{T_i^-(k)}}{1+\sum_{k=1}^N \prod_{\ell=1}^k \frac{T_i^+(\ell-1)}{T_i^-(\ell)}}.
\end{equation}
This expression for the marginal stationary distribution can be obtained by standard methods from the backward master equation (see for example \cite{Ewens2004,Traulsen2009}). We will test these predictions against simulations for the homogeneous and heterogeneous noisy voter models in Sections~\ref{sec:hom} and~\ref{sec:het}. In the figures and in Appendix \ref {sec:diffapprox} we express the marginals of the stationary distribution in terms of ${x_i=n_i/N}$; we then use the notation ${{\cal P}_i(x_i)}$.

\subsection{Noise-induced transition}\label{sec:nit}

As discussed above, noisy voter models show a transition between parameter ranges with unimodal and multi-modal stationary distributions. This transition is noise induced, and occurs as the system size is varied. As a consequence, it is useful to identify a critical population size $N_c$ which separates the two types of outcomes.  In the case of only two species ($m=2$) this is particularly straightforward, as there exists a population size for which the stationary distribution becomes flat. This population size has been used to define the transition point in refs.~\cite{Biancalani2014,Houchmandzadeh2015}; see also \cite{Carro2016}.

In the multi-state case ($m>2$) the situation is more complicated; the marginal distributions for the $\{x_i\}$ do not assume a flat shape for any population size, as we will discuss below. In order to characterise the transition, we therefore consider the shape of the distribution $P_i(n_i)$ at the left and right boundaries of phase space, i.e., near $n_i=0$ and $n_i=N$, respectively. Specifically we find the population sizes $N_L$ and $N_R$ for which $P_i(0)=P_i(1)$ and $P_i(N-1)=P_i(N)$, respectively. These can be used to separate the unimodal and multi-modal regimes. We will discuss this in more detail for the homogeneous and heterogeneous multi-state models in Secs.~\ref{sec:hom} and \ref{sec:het}.

\subsection{Switching times between states of consensus}\label{sec:SwitchTimes}
The system is said to have reached consensus on opinion $i$ if $n_i=N$, implying $n_j=0$ for all $j\neq i$. We refer to this as consensus state $i$. If the population is not in a consensus state, then at least two of the $n_j$ are non-zero, i.e., several opinions are represented in the population; we call these situations mixed states. As shown in Fig.~\ref{fig:JustVote}, the system can transition between consensus states, with intermediate periods in mixed states. We will refer to this as `switching' between consensus states, and we proceed to calculate the typical time between such switches.

\begin{figure}[b!]
	\center
	\includegraphics[width=0.4\textwidth]{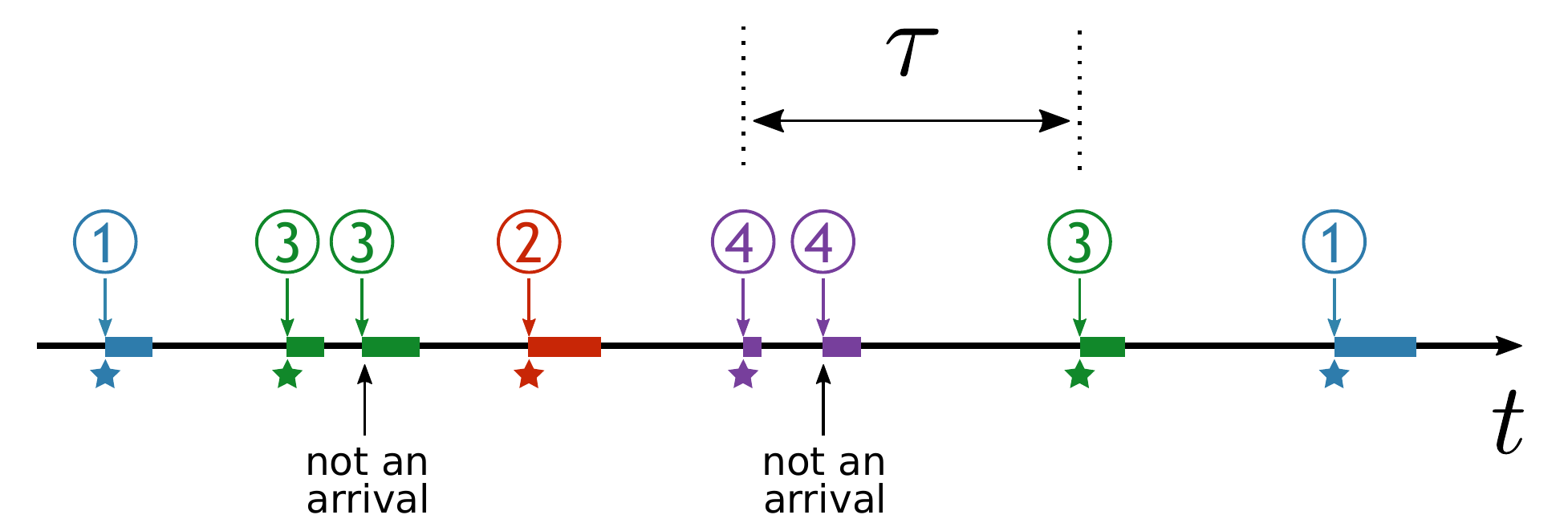}
	\caption{\textbf{Illustration of the concepts of arrival and switching time.}
		We show the time line of a model with $m\geq 4$ opinion states. Times at which the system reaches a consensus state are marked above the time axis by circled numbers. Times during which the system resides at a consensus state are indicated as filled bars on the time axis. Between these times the population is in mixed states. Arrivals at a new consensus state, as defined in the text, are marked by stars below the time axis. The switching time $\tau$ is the mean time between subsequent arrivals at new consensus states.
		}
	\label{fig:arrivals}
\end{figure}

In order to define switching times, we first  introduce the concept of an `arrival' at a consensus state. We say that an arrival at consensus state $i$ occurs at time $t$ if the system transitions from a mixed state into consensus state~$i$ at time~$t$, {\em and} if the last consensus state the population visited before $t$ was not $i$. This is illustrated in Fig.~\ref{fig:arrivals}. 

We define the mean switching time $\tau$ as the average time that elapses between two subsequent arrivals. By definition, these arrivals are at two different consensus states. This is again illustrated in Fig.~\ref{fig:arrivals}. The quantity $\tau^{-1}$ is the mean number of arrivals per unit time. 

In order to compute $\tau$, we generalized the procedure used in ref.~\cite{Houchmandzadeh2015}. First, we consider the mean time it takes a single species $j$ to reach state $n_j=N$ starting from $n_j=0$; we denote this time by $t_j^{0\to N}$. This is a standard hitting-time problem, and we find \cite{Norris1998,Traulsen2009}
	\begin{equation}\label{eq:t0N}
	t_j^{0\to N}=\sum_{k=0}^{N-1}\sum_{\ell=0}^k \frac{1}{T_j^+(\ell)}\prod_{\ell'=\ell+1}^k \frac{T_j^-(\ell')}{T_j^+(\ell')}.
	\end{equation}

To proceed, we now assume that the population is currently in consensus state $i$. The mean time it takes any of the 
species $j\neq i$ to reach the state $n_j=N$ is then given by $t_j^{0\to N}$. This expression is exact, provided closed-form processes can be formulated for the individual species. 

To be able to describe the switching time of the multi-state model, we now treat the processes for each of the species $j\neq i$ as independent. We associate each process with a `clock', and say that the clock for species $j$ `ticks' when $n_j=N$. We assume that the rate with which this happens is constant in time for each process, and given by $1/t_j^{0\to N}$. This is an approximation. First, due to the coupling of the different species, the processes for the different types $j$ are in general not independent. Second, the distribution of waiting times until state $n_j=N$ is reached is not exponential for one-step Markov processes of the type above; instead, the hitting time can be written as the sum of multiple exponential random variables \cite{Karlin1959, Keilson2012markov}. 

Proceeding nevertheless with this approximation, the total rate for any of the clocks $j\neq i$ to tick (i.e., any species $j\neq i$ to take over the population) is given by $\sum_{j\neq i} (t_j^{0\to N})^{-1}$. As a consequence, we find the mean waiting time until any of the consensus states $j\neq i$ is reached as
	\begin{equation}\label{eq:tau_i}
	\tau_i=\frac{1}{\sum_{j\neq i}  (t_j^{0\to N})^{-1}}.
	\end{equation}
This approximates the average time for the system to reach any consensus state $j\neq i$, if it is currently in consensus state $i$. 

The mean time between two subsequent arrivals can then be written as
	\begin{equation}\label{eq:tau}
	\tau=\sum_{i=1}^m p_i \tau_i,
	\end{equation}
where $p_i$ denotes the proportion of the number of arrivals at consensus state $i$ among the total number of arrivals. In other words, of all arrivals the system makes, a fraction $p_i$ occurs at consensus state~$i$. Evidently, one has ${\sum_{i=1}^m p_i=1}$.  

We now proceed to estimate the coefficients $p_i$. To do this, we first consider the typical time between two arrivals at the same consensus state $i$. This can be approximated as
	\begin{equation}\label{eq:tauret}
	\overset{\text{\scriptsize$\leftrightarrow$}}{\tau}_i\equiv \tau_i+t_i^{0\to N}.
	\end{equation}
This expression can be understood by first assuming that the population is in consensus state $i$.  The term $\tau_i$ in Eq.~(\ref{eq:tauret}) is the mean time that elapses until the system reaches any of the other consensus states, $j\neq i$. At that point there will be no individuals of type $i$ ($n_i=0$). The second term, $t_i^{0\to N}$, is the average time required to reach consensus state $i$ again.

The quantity $p_i$ in turn is proportional to the number of arrivals at $i$ per unit time, given by $(\overset{\text{\scriptsize$\leftrightarrow$}}{\tau_i})^{-1}$. 
With appropriate normalization, we find
	\begin{equation}\label{eq:p_i}
	p_i=\frac{ (\overset{\text{\scriptsize$\leftrightarrow$}}{\tau}_i)^{-1}}{\sum_{j=1}^m  (\overset{\text{\scriptsize$\leftrightarrow$}}{\tau}_j)^{-1}  }.
	\end{equation}
	
To summarise, the overall switching time is obtained as
	\begin{equation}\label{eq:taufinal}
	\tau=\sum_i \left\{\frac{ (\overset{\text{\scriptsize$\leftrightarrow$}}{\tau}_i)^{-1}}{\left[\sum_{j}  (\overset{\text{\scriptsize$\leftrightarrow$}}{\tau}_j)^{-1}\right]\left[ \sum_{j\neq i}  (t_j^{0\to N})^{-1}\right]}\right\},
	\end{equation}
using the expressions in Eqs.~(\ref{eq:t0N}), (\ref{eq:tau_i}), and (\ref{eq:tauret}).

\section{Homogeneous model}\label{sec:hom}
We first consider the fully homogeneous set-up. If the rates are homogeneous across species, i.e.,  if $r_{ij}\equiv r$ and $\varepsilon_{ij}\equiv \varepsilon$, the expressions in Eq.~(\ref{eq:tpm}) reduce to
	\begin{align}\label{eq:homrates}
	T_i^+(n_i)=& \frac{n_i (N-n_i)}{N}r+(N-n_i)\varepsilon, \nn\\
	T_i^-(n_i)=&\frac{n_i(N-n_i)}{N}r+\varepsilon (m-1) n_i.
	\end{align}
In particular, the $T_i^\pm$ are now fully specified by $n_i$ alone. Species $i$ undergoes a well-defined death-birth process with rates given by Eq.~(\ref{eq:homrates}). All species $j\neq i$, can be lumped together into one `other species' and there are $N-n_i$ individuals of this other species.  We note that no approximations have been made in deriving this result. As far as the marginal process for any particular species $i$ is concerned, the dynamics effectively reduces to a two-species process, for which analytical descriptions are available (see e.g. \cite{Ewens2004}). We will use these tools to calculate stationary distributions for individual species, as well as switching times between consensus states.

\subsection{Marginals of the stationary distribution}
\begin{figure*}[t!]
	\center
	\includegraphics[width=0.8\textwidth]{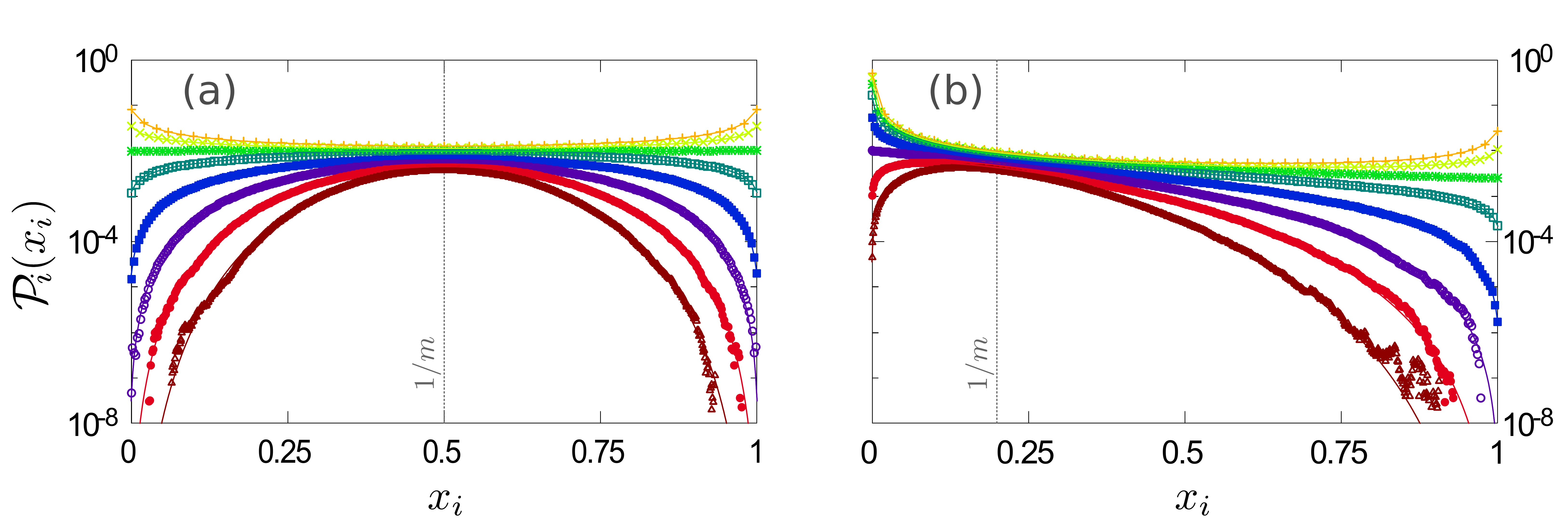}
	\caption{\textbf{Stationary distribution of the model with homogeneous rates across species.}
		Panel (a) is for $m=2$; panel (b) shows the marginal distribution for single species for the model with $m=5$. The different curves are for different population sizes in the range $N=50$ (top) to $N=900$ (bottom). Markers are from simulations; lines show the analytical predicitions from the theory described in the text. Remaining model parameters are $r=1$, and $\epsilon=10^{-2}/(m-1)$.}
	\label{fig:ProbState}
\end{figure*}
For uniform rates, the marginal probability distributions $P_i(n_i)$ are identical across species and can be obtained by using the rates from Eqs.~(\ref{eq:homrates}) in Eq.~(\ref{eq:statp}). These closed-form expressions can be evaluated numerically.

Results can be found in Fig.~\ref{fig:ProbState}. In panel (a) we show the case of $m=2$ species, already discussed in refs.~\cite{Biancalani2014,Houchmandzadeh2015}. This is shown mainly for comparison.

Given that ${n_2=N-n_1}$, the model with two species is, by construction, described by a single death-birth process. As seen in the figure, simulation results and theory are in agreement. For $N>N_c$ the resulting distribution is unimodal; for $N<N_c$ a bimodal shape is observed. At the critical population size $N=N_c$ the distribution is flat; this transition point is $N_c=r/\varepsilon$ \cite{Biancalani2014,Houchmandzadeh2015}. The stationary distribution is symmetric about $x=1/2$ for all $N$.

In panel (b) of Fig.~\ref{fig:ProbState} we show the model with $m=5$ opinion states. Again, simulations and theoretical predictions are in agreement. As in the case of two states, a transition from a unimodal to a bimodal marginal distribution is observed as the population size is decreased. Bimodality in the marginal single-species distributions indicates multi-modality in the state space of all $m$ species, as illustrated in Fig.~\ref{fig:JustVote}(c) for $m=3$. In contrast with the two-species case, the marginal distributions for the abundances of individual species never becomes flat in Fig.~\ref{fig:ProbState}(b). We also note that the marginal distributions do not exhibit any particular symmetry, despite the fact that the imitation and mutation rates are homogeneous across species.

\subsection{Phase diagram}\label{sec:pghom}
As mentioned in Sec.~\ref{sec:nit}, we use the population sizes at which ${P_i(0)=P_i(1)}$ or ${P_i(N-1)=P_i(N)}$ to characterise the transition between the unimodal and multi-modal regimes.
The condition for the left edge, $P_i(0)=P_i(1)$, translates into ${T_i^+(0)=T_i^-(1)}$; using Eqs.~(\ref{eq:homrates}) one finds the physical solution

\begin{equation}\label{eq:nl}
	N_L=\frac{\left[1+\widetilde\varepsilon_{\rm tot} \right]+\sqrt{\left[1+\widetilde\varepsilon_{\rm tot}\right]^{2}-4\widetilde\varepsilon}}{2\widetilde\varepsilon},
	\end{equation}
where we have written $\widetilde\varepsilon_{\rm tot}=(m-1)\varepsilon/r$, and $\widetilde\varepsilon=\varepsilon/r$.

Similarly, requiring $P_i(N-1)=P_i(N)$ at the right edge yields $T_i^+(N-1)=T^-(N)$, from which we obtain

\begin{equation}\label{eq:nr}
	N_R=\frac{\left[1+\widetilde\varepsilon \right]+\sqrt{\left[1+\widetilde\varepsilon\right]^{2}-4\widetilde\varepsilon_{\rm tot}}}{2\widetilde\varepsilon_{\rm tot}}.
	\end{equation}
For $m=2$ this reduces to the result in refs.~\cite{Biancalani2014,Houchmandzadeh2015}, i.e., ${N_L=N_R=r/\varepsilon}$.

The resulting expressions for $N_L$ and $N_R$ only depend on the relative mutation rate $\varepsilon/r$. This is due to the fact that, up to a re-scaling of time, the reaction rates in Eq.~(\ref{eq:homrates}) only depend on the ratio of $\varepsilon$ and $r$. We also note that Eqs.~(\ref{eq:nl}) and (\ref{eq:nr}) are symmetric with respect to exchanging $\widetilde\varepsilon$ and $\widetilde\varepsilon_{\rm tot}$. This is a consequence of the same symmetry in the pairs $\{T^+(0), T^-(N)\}$  and $\{T^-(1), T^+(N-1)\}$. We also notice that ${N_L\approx \widetilde\varepsilon^{-1}}$ and ${N_R\approx (\widetilde\varepsilon_{\rm tot})^{-1}}$ when ${\varepsilon/r \ll 1}$.

\begin{figure*}[t!]
	\center
	\includegraphics[width=0.8\textwidth]{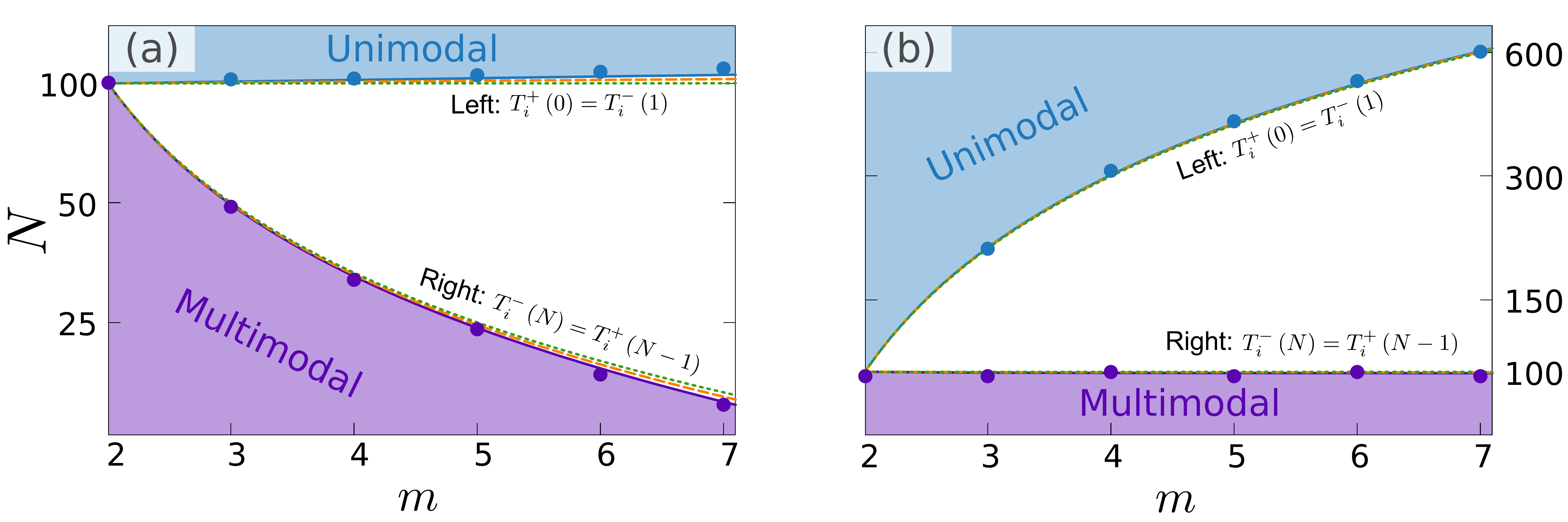}
	\caption{\textbf{Phase diagram for the model with homogeneous rates.}
	The critical system size is plotted as a function of the number of strategies.
	The continuous blue line is $N_L$ as calculated from Eq.~(\ref{eq:nl}), and the purple continuous line shows $N_R$ as obtained from Eq.~(\ref{eq:nr}). The remaining lines are from a diffusion approximation to the model, as discussed in Appendix \ref{sec:diffapprox}. Markers show results from simulations. Mutation rates are $\varepsilon=10^{-2}$ in panel (a), and $\varepsilon=10^{-2}/(m-1)$ in panel (b). We set $r=1$ in both panels.}
	\label{fig:CriticalN}
	\end{figure*}

The phase diagram resulting from Eqs.~(\ref{eq:nl})~and~(\ref{eq:nr}) is depicted in Fig.~\ref{fig:CriticalN}. We use $r=1$ in both panels. In panel (a) we show the population sizes $N_R$ and $N_L$ as a function of the number of species, $m$, at a fixed mutation rate $\varepsilon$.  The upper sold line indicates $N_L$ and the lower solid line is $N_R$. Results from simulations are shown as markers; as seen in the figure, theoretical predictions are in agreement with simulations. 

We note that, in Fig.~\ref{fig:CriticalN}(a), $N_L$ is nearly flat as a function of the number of species $m$. This is in line with the behaviour of $N_L\approx r/\varepsilon$ for small mutation rates, and can be further understood as follows. The shape of the stationary state distribution near $n_i=0$ is largely determined by the frequency of mutation events that generate new individuals of type $i$. This mutation rate is given by $N\varepsilon$; if $n_i=0$, each of the $N$ individuals in the population mutate into $i$ with rate $\varepsilon$ [see Eq.~(\ref{eq:homrates})]. This is the case independently of the number of species $m$. Therefore, we expect $N_L$ to be approximately independent of $m$, as seen in Fig.~\ref{fig:CriticalN}(a). 

Near $n_i=N$, however, the shape of the distribution is determined by the rate with which individuals of type $i$ mutate into any other species. If $n_i=N$, each of the $N$ individuals mutates into any one of the species $j\neq i$ with rate $\varepsilon$; the total rate of mutation events out of species $i$ is then $N\varepsilon(m-1)$ [see again Eq.~(\ref{eq:homrates})]. For fixed $\varepsilon$, this total rate increases as either $N$ or $m-1$ are increased. Therefore, we expect $N_R$ to be a decreasing function of $m$, which is confirmed in the phase diagram in Fig.~\ref{fig:CriticalN}(a). This is again consistent with the behaviour $N_R\approx r/[(m-1)\varepsilon]$ for small mutation rates.

In panel (b) of Fig.~\ref{fig:CriticalN} we show the phase diagram for the choice $\varepsilon=10^{-2}/(m-1)$, i.e., for a fixed value of $\widetilde\varepsilon_{\rm tot}=10^{-2}$. Recalling that an individual of any species undergoes mutations to any of the other $m-1$ species with rate $\varepsilon$, the {\em total} mutation rate for any individual is constant with this choice, irrespective of the number of species $m$. The model shows the same three phases as in Fig.~\ref{fig:CriticalN}(a), but the shape of the phase lines changes. We now find that $N_L$ is nearly independent of $m$, whereas $N_R$ is a decreasing function of $m$.  This is consistent with the interpretation given above. The shape of the distribution near the left edge is largely determined by ${T_i^+(0)=10^{-2}N/(m-1)}$. An increased value of $m$ now requires a larger value of $N$ to keep this rate constant. The shape of the distribution on the right edge is governed by ${T_i^-(N)=10^{-2}N}$; this quantity is now independent of the number of species, $m$. 

Similar to \cite{Houchmandzadeh2015}, our approach so far has not involved any continuous approximation of the discrete state space of the population. The analysis of ref.~\cite{Biancalani2014}, on the other hand, is based on a diffusion approximation to the dynamics of the two-state noisy voter model. We have extended this approach to the multi-state case. The dashed and dotted lines in both panels of Fig.~\ref{fig:CriticalN} are the predictions for $N_L$ and $N_R$ obtained from the diffusion approximation. The mathematical details are discussed in the Appendix.

The shape of $P_i(n_i)$ in the different regimes of Fig.~\ref{fig:CriticalN} can be summarised as follows [see also Fig.~\ref{fig:ProbState}(b)]: For large populations (${N>N_L}$) the single-species distribution is unimodal (light blue shading in Fig.~\ref{fig:CriticalN}), taking its maxi-mum strictly in the interior of the interval ${0<n_i<N}$. For intermediate sizes (${N_R<N<N_L}$, unshaded) the distribution function is monotonously decreasing in $n_i$, taking its maximum value at ${n_i=0}$. Finally, for small populations (${N<N_R}$) the distribution for $n_i$ is multi-modal (dark purple shading), with maxima at ${n_i=0}$ and ${n_i=N}$.

\subsection{Switching times}
In the homogeneous case, the times $t_j^{0\to N}$ in Eq.~(\ref{eq:t0N}) are uniform across species; we write $t^{0\to N}$ for their common value. As a consequence, we find $\tau_i=t^{0\to N}/(m-1)$ in Eq.~(\ref{eq:tau_i}), and therefore the mean switching time is also given by 
	\begin{equation}\label{eq:tshom}
	\tau=\frac{t^{0\to N}}{m-1}.
	\end{equation}

\begin{figure}[t!]
	\center
	\includegraphics[width=0.45\textwidth]{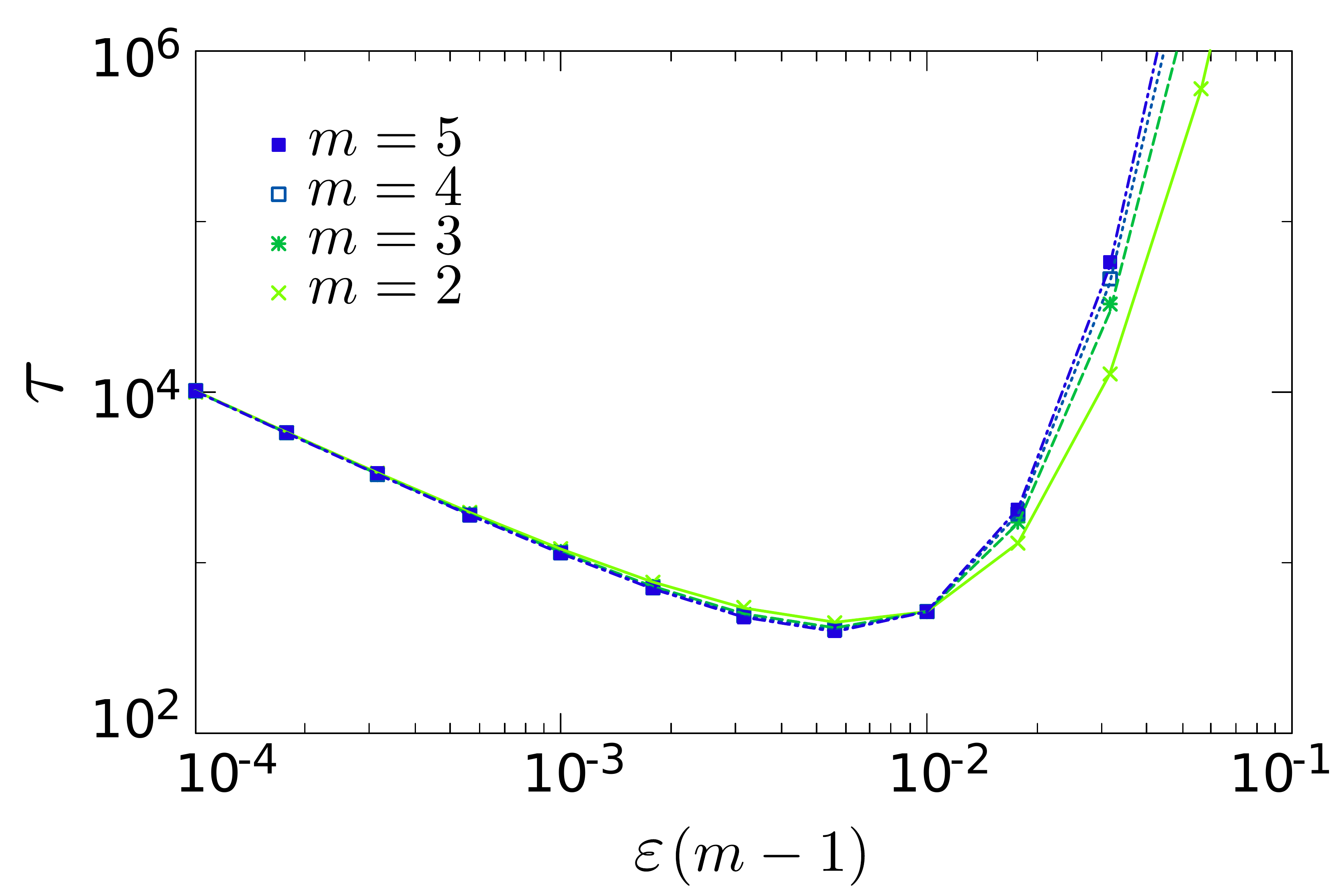}
	\caption{\textbf{Switching times in the model with homogeneous imitation and mutation rates.}
		The figure shows the switching time $\tau$ between consensus states for different choices of the number of species $m$. Lines are from Eq.~(\ref{eq:tshom}), markers show simulation results. In all cases $N=100$ and $r=1$. 
	}
	\label{fig:SwitchTimesHom}
\end{figure}

Theoretical predictions and results from simulations are compared in Fig.~\ref{fig:SwitchTimesHom}. We show data for different choices of the number of species $m$; as before we use $r=1$. To allow a better comparison, we plot the switching time as a function of $\widetilde\varepsilon_{\rm tot}=(m-1)\varepsilon$. 
This ensures that the total mutation rate any one species experiences is comparable across the different values of $m$. 

The diagram indicates that the switching time becomes smallest at intermediate values of the mutation rate. This can be understood as follows. For small mutation rates, the population will reside at consensus states for relatively long times; escape from these states occurs through mutation. These long sojourn times at the consensus state mean that the switching time, i.e., the typical time between arrivals, will also be large. On the other hand, for large mutation rates the population spends most of its time in the interior of phase space. This leads to long periods in which the population does not visit any of the consensus states. As a consequence, the switching time will also be large. The shortest switching times are thus seen in the intermediate range of mutation rates, when neither of these effects dominate.

\section{Heterogeneous model}\label{sec:het}
We now turn to instances of the model in which the imitation and mutation rates are not homogeneous across species. We limit the discussion to examples in which $r_{ji}=r_j$ and $\varepsilon_{ji}=\varepsilon_j$. The approach we develop can be extended to cover more general cases; we briefly comment on this in Appendix \ref{sec:genhet}.

The quantity $r_j$ describes the rate with which members of species $j$ imitate any other species, and $\varepsilon_j$ is the rate with which individuals of type $j$ spontaneously change to  any other type. In the context of opinion dynamics, large values of $r_j$ and $\varepsilon_j$ thus describe beliefs that are only weakly held. If $r_j$ and $\varepsilon_j$ are low, on the other hand, then $j$ describes a strong view. 

\subsection{Analytical approximation}\label{sec:aapprox}
With the above choice of reaction rates, the transition rates $T_i^\pm$ in Eq.~(\ref{eq:tpm}) can be written as
	\begin{align}\label{eq:hetrates}
	T_i^+(\bn)&= \frac{n_i (N-n_i)}{N} \avg{r}_{-i}(\bn)+(N-n_i) \avg{\varepsilon}_{-i}(\bn), \nn\\
	T_i^-(\bn)&= \frac{n_i(N-n_i)}{N}r_i+\varepsilon_i n_i,
	\end{align}
where
	\begin{equation}\label{eq:r*-i}
	\avg{r}_{-i}(\bn)=\frac{1}{N-n_i}\sum_{j\neq i} n_j r_j,
	\end{equation}
and similarly for $\avg{\varepsilon}_{-i}(\bn)$. These quantities are weighted averages over the imitation and mutation rates of all species other than $i$; $\avg{r}_{-i}$ for example represents the mean rate with which opinion $i$ is imitated by any other species. We note that  $\avg{r}_{-i}(\bn)$ and  $\avg{\varepsilon}_{-i}(\bn)$ depend on the variables $n_j$ ($j\neq i$), i.e., on the entire state of the population. As a consequence, no closed process can be formulated for individual species for the heterogeneous model. 

We therefore resort to an approximation. We replace all $\{n_j\}$ with $\{N x_j^*\}$ in Eq.~(\ref{eq:r*-i}); $x_j^*$ represents the proportion of individuals of species $j$ at the fixed point of the deterministic rate equations, i.e., the dynamics in the limit of infinite populations. These can be obtained from a Kramers--Moyal expansion of the master equation~(\ref{eq:master}), as detailed in Appendix \ref{sec:diffapprox}. We then define
	\begin{equation}\label{eq:rstar}
	r_{-i}^*\equiv\frac{1}{1-x_i^*}\sum_{j\neq i} x_j^* r_j,
	\end{equation}
and similarly for $\varepsilon^*_{-i}$.
Our approximation consists in replacing the stochastic quantities $\avg{r}_{-i}(\bn)$ and $\avg{\varepsilon}_{-i}(\bn)$ with $r_{-i}^*$ and $\varepsilon^*_{-i}$, respectively, in Eq.~(\ref{eq:hetrates}). In this way we find an approximate, but closed, effective process for species $i$. This can then be used to obtain marginals of the stationary distribution and switching times, following the procedure described in Sections~\ref{sec:StatDist}~and~\ref{sec:SwitchTimes}.

\subsection{Marginals of the stationary distribution}\label{sec:mhet}
In order to test the accuracy of the approximations described above, we consider an example with $m=5$ species. We choose the imitation rates as ${r_j=r[1-\delta+(j-1)\frac{2\delta}{m-1}]}$, for $j=1,\dots,m$, and simi-larly ${\varepsilon_j=\varepsilon[1-\delta+(j-1)\frac{2\delta}{m-1}]}$ for the mutation rates. The coefficients $\varepsilon$ and $r$ represent the mean rates (across species) and are model parameters.
This choice indicates that the rates $\{r_j\}$ and $\{\varepsilon_j\}$ are equally spaced in the intervals $[(1-\delta)r,(1+\delta)r]$ and $[(1-\delta)\varepsilon,(1+\delta)\varepsilon]$, respectively. The parameter $\delta$ therefore describes the relative spread of these rates, and quantifies the degree of heterogeneity. We note that species $j=1$ has the lowest rates $\varepsilon_j$ and $r_j$ among all species (it represents a strong opinion), and species $j=m$ the highest (it describes a weak opinion).
We restrict our analysis to relatively small values of $\delta$; this avoids situations in which the population spends most of its time in one single consensus state (the one of the strongest opinion). In such cases, it is difficult to measure stationary distributions and switching times in simulations.

\begin{figure*}[t!]
	\center
	\includegraphics[width=0.95\textwidth]{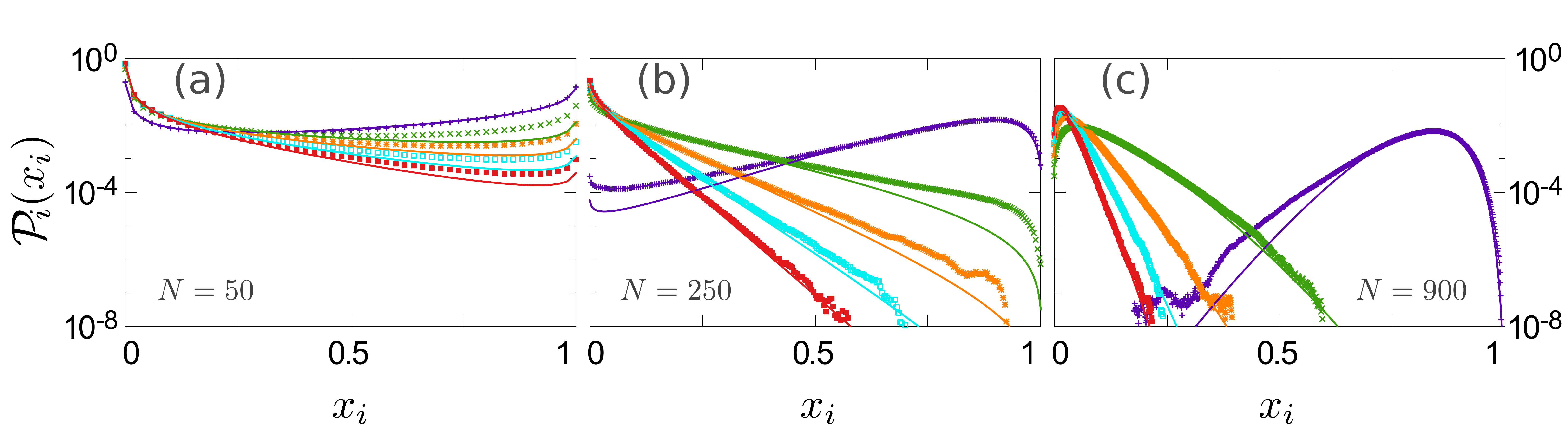}
	\caption{\textbf{Marginals of the stationary distribution for the heterogeneous multi-state noisy voter model.}
	We show the distributions ${\cal P}_i(x_i)$ for the different individual species in a model with $m=5$ for different population sizes, as indicated in the figure. Markers represent simulation results; lines are evaluations of Eq.~(\ref{eq:statp}), using the rates in Eq.~(\ref{eq:hetrates}) with the approximation in Eq.~(\ref{eq:rstar}). The choice of mutation and imitation rates is as described in the text (see Sec.~\ref{sec:mhet}), using $r=1$, $(m-1)\varepsilon=10^{-2}$, and $\delta=0.05$.}
	\label{fig:StatP_het}
	\end{figure*}

The marginals of the resulting stationary distribution are shown in Fig.~\ref{fig:StatP_het}.  In panel (a) the population is relatively small ($N=50$), panel (b) shows an intermediate case ($N=250$), and in panel (c) $N=900$. 
As the population size is increased, the marginal distributions change shape. Each marginal is bimodal for very small populations [panel (a)]; in particular, the slope of each curve is negative near $n_i=0$, and positive near $n_i=N$. For larger populations the gradient near $n_i=N$ changes sign when a critical size, $N_{R,i}$, is reached [panel (b)], similar to what was found in the homogeneous model. Notably though the population size at which this happens can differ across the species, as indicated by the subscript $i$ in $N_{R,i}$. If the population size is larger still, the gradient of the marginal distribution near $n_i=0$ will also change sign [panel (c)]; this occurs at population sizes $N_{L,i}$, which can again vary across the different species. Mathematically, $N_{L,i}$ and $N_{R,i}$ can be defined using conditions similar to those in Sec.~\ref{sec:pghom}; $N_{L,i}$ is the population size for which $T_i^+(0)=T_i^-(1)$, and $N_{R,i}$ is defined as the population size for which $T_i^+(N-1)=T_i^-(N)$.

\subsection{Phase diagram}

In Fig.~\ref{fig:StatP_het}(a) the population size is such that $N<N_{R,i}$ for all $i$, so that all marginals have bimodal shape. In panel (b) we have $N_{R,i}<N<N_{L,i}$ for all $i$, and in panel (c) the population is sufficiently large so that $N>N_{L,i}$ for all $i$.

\begin{figure}[b!]
	\center
	\includegraphics[width=0.41\textwidth]{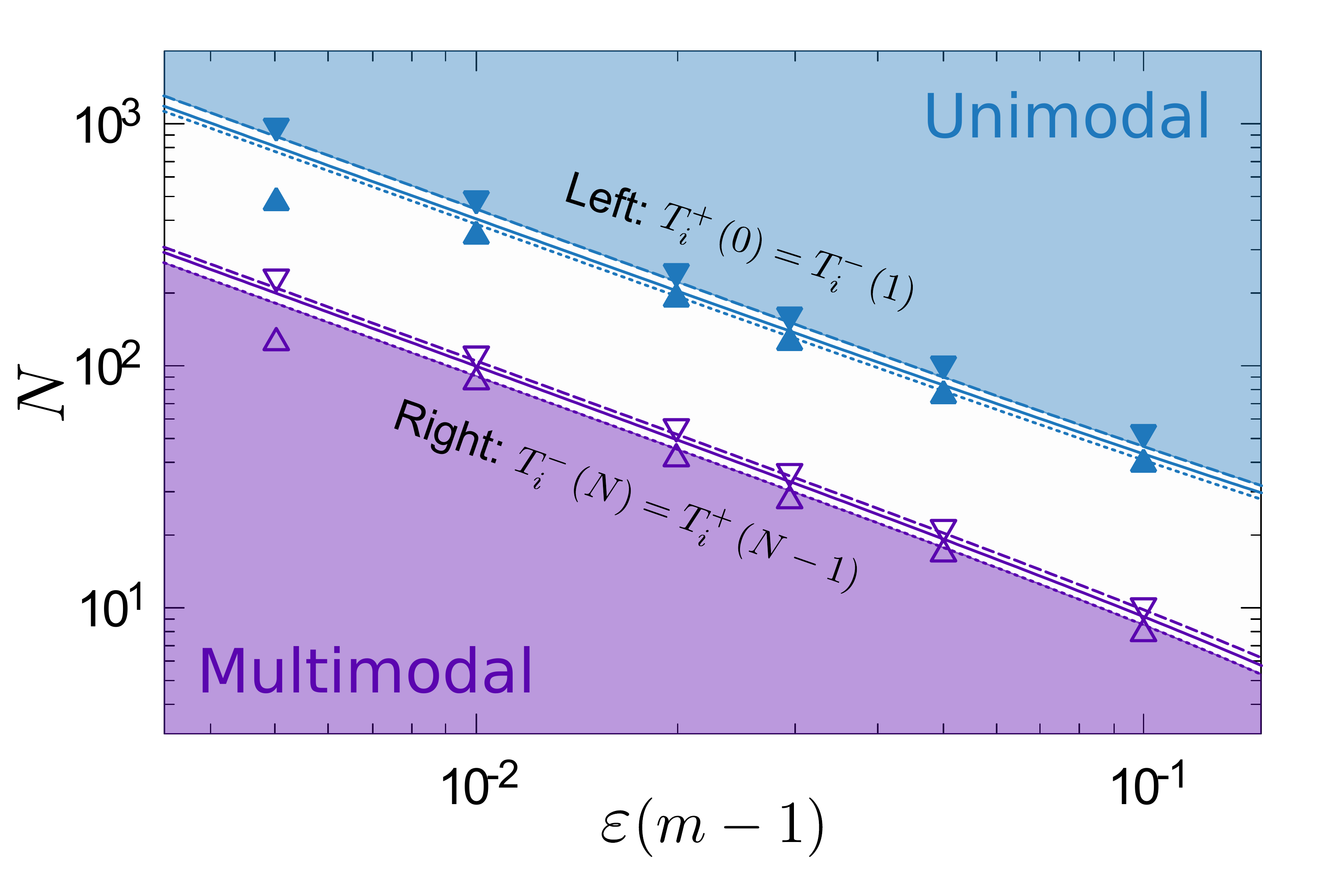}
	\caption{\textbf{Phase diagram of the model with $m=5$ species and heterogeneous rates.}
	Upper and lower dashed lines show $N_L^{\rm max}$ and $N_R^{\rm max}$, respectively; upper and lower dotted lines are $N_L^{\rm min}$ and $N_R^{\rm min}$. Solid lines are $N_L^{\rm hom}$ and $N_R^{\rm hom}$ (see text for definitions). Markers are from simulations (with $\blacktriangledown, \blacktriangle, \triangledown, \vartriangle$ showing $N_L^{\rm max}, N_L^{\rm min}, N_R^{\rm max}, N_R^{\rm min}$, respectively). Mutation and imitation rates for each species are chosen as described in the text (see Sec.~\ref{sec:mhet}).}
		\label{fig:phasediagram_het}
\end{figure}
The resulting phase diagram for the model with heterogeneous rates is shown in Fig.~\ref{fig:phasediagram_het}. It exhibits the three different phases outlined above. 

For a given value of the mean mutation rate, the marginal distributions for all species are unimodal if $N>N_{L,i}$ for all $i$. 
On the other hand, if $N<N_{R,i}$ for all $i$ then all marginals are bimodal. It is useful to define $N_L^{\rm min}\equiv \mbox{min}_i\,N_{L,i}$, and $N_L^{\rm max}\equiv\mbox{max}_i\, N_{L,i}$, and similarly for $N_R^{\rm min}$ and $N_R^{\rm max}$. We also define $N_{L}^{\rm hom}$ and $N_{R}^{\rm hom}$ as the corresponding quantities for the homogeneous model, with $r_i\equiv r, \varepsilon_i\equiv \varepsilon$ for all $i$ [see Eqs.~(\ref{eq:nl}) and (\ref{eq:nr})].
 
In Fig.~\ref{fig:phasediagram_het}, we plot these quantities for different choices of $m$. The white unshaded region shows population sizes for which $N_R^{\rm min}<N<N_L^{\rm max}$. Simulation results for $N_R^{\rm min}$, $N_R^{\rm max}$, $N_L^{\rm min}$, and $N_L^{\rm max}$ are also shown, and confirm the theoretical predictions.

\subsection{Switching times}
\begin{figure*}[t!]
	\center
	\includegraphics[width=0.9\textwidth]{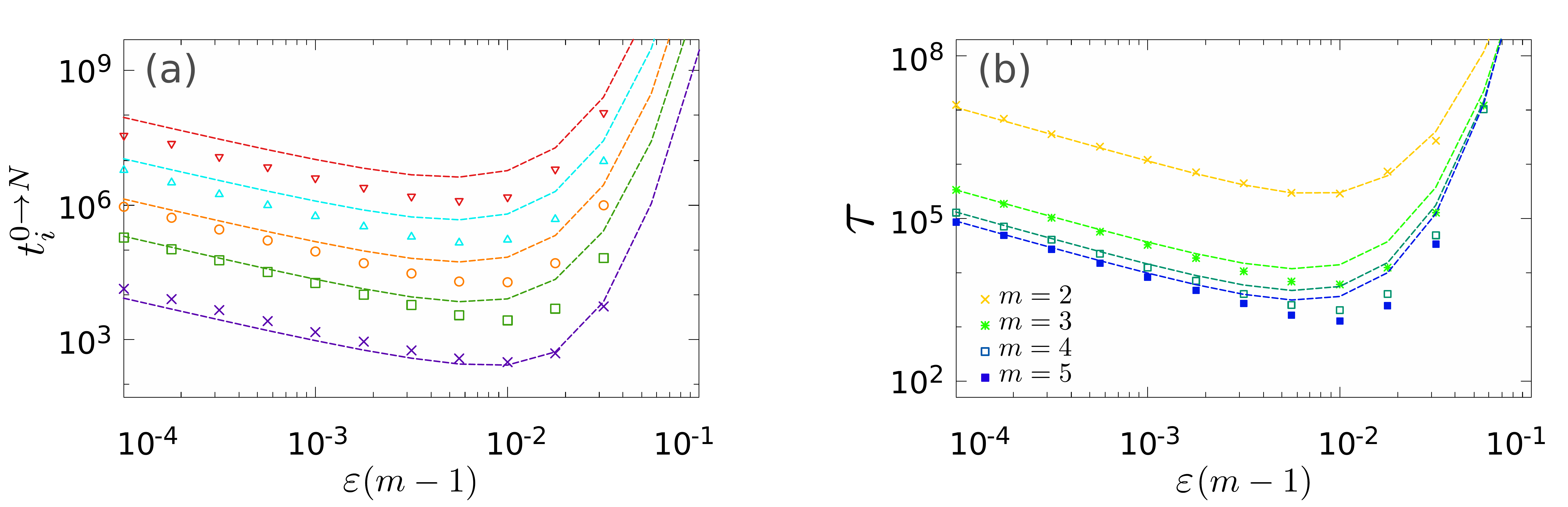}
	\caption{\textbf{Switching times in the noisy voter model with heterogeneous rates.}
		Panel (a) shows the quantities $t_i^{0\to N}$ for the different species $i=1,\dots, 5$ (bottom to top) in the model with $m=5$. Panel (b) shows the resulting switching time $\tau$ for models with $m=2,3,4,5$ species, from top to bottom. Lines are from the analytical approximation [Eq.~(\ref{eq:taufinal}) with the rates as approximated in Eq.~(\ref{eq:rstar})]; markers are from simulations. In all cases, $N=100$, $\delta=0.05$; imitation and mutation rates are distributed as described in Sec.~\ref{sec:mhet}, using $r=1$ and $\varepsilon$ as indicated on the horizontal axis.
		}
		\label{fig:SwitchTimesHet}
\end{figure*}

In Fig.~\ref{fig:SwitchTimesHet} we demonstrate the accuracy of the analytical approximation in Eq.~(\ref{eq:taufinal}) for the switching times in the model with heterogeneous rates. Panel (a) shows the times $t_i^{0\to N}$ for the different species $i$ in the model with $m=5$. Imitation and mutation rates were chosen as described in Sec.~\ref{sec:mhet}. As seen in the figure, the overall agreement is reasonable, although not quantitatively accurate. This is not surprising, as the approximation of  Eq.~(\ref{eq:rstar}) is a relatively severe intervention. Nevertheless, the approximation accurately predicts the qualitative shape of $t_i^{0\to N}$ as a function of the mutation rate, as well as the ordering across species. We note that the different curves in panel (a) show species $i=1,\dots,5$ from bottom to top. This is in-line with intuition: individuals who are of opinion $i=5$ are easily convinced of other opinions and are likely to change their views spontaneously. It is therefore difficult for this opinion to spread in the population; as a consequence $t_5^{0\to N}$ is large. Individuals who are of opinion $i=1$, on the other hand, do not easily change to any other opinion; this leads to a relatively small time $t_1^{0\to N}$.

In Fig.~\ref{fig:SwitchTimesHet}(b) we show the resulting switching time $\tau$ for the model with heterogeneous rates. Results are reported for models with different numbers of species, $m$, and as a function of $(m-1)\varepsilon$ as in previous figures. In each case the switching time shows a minimum as a function of the mean mutation rate, as in the homogenous model.

It is interesting to note that the switching time is found to decrease at a fixed mutation rate $(m-1)\varepsilon$ when the number of species increases. 
To understand this, we observe that the spacing between rates for different species is proportional to $\delta/(m-1)$; see the definitions at the beginning of Sec.~\ref{sec:mhet}. At fixed $\delta$ species $i=1$ is thus the strongest opinion by a large margin when $m$ is small. 
In the extreme case there are only two opinions; one is a very strong view and the other very weak in comparison. The population visits consensus on the stronger opinion frequently, and remains there for relatively long times. This leads to long switching times. When there are more species available (larger values of $m$) and $\delta$ is kept fixed, the opinion spectrum becomes more finely spaced and the advantage of the most strongly held view is reduced. This leads to a reduction in switching times. 

A similar effect occurs when the parameter $\delta$ is varied keeping all other model parameters fixed. This parameter controls the spread of the rates $r_j$ and $\varepsilon_j$. We find in simulations that switching times increase with increasing spread~$\delta$. This is again due to an increased advantage of the most strongly held opinion.

\section{Summary and discussion}\label{sec:concl}
We have introduced a multi-state noisy voter model and studied its behaviour in a population with all-to-all interaction. It can be seen as a stylised model of imitation and mutation processes in opinion dynamics or evolution, for example. The model shows a noise-driven transition between a state of diversity for large populations and a mostly ordered state in small populations. In the diverse state multiple opinions (or species) are present in the population at most times, and the resulting stationary distribution is unimodal, with a peak in the interior of state space. The ordered regime occurs for small population sizes; the system is frequently found in consensus states, in which all individuals are of the same type. Switches between these consensus states are possible, as the dynamics do not permit absorbing states for non-zero mutation rates.

The transition between the two types of behaviour is a consequence of the balance of mutation effects, driving the system towards mixed states, and intrinsic noise due to the discreteness of the population. Similar to what was reported in ref.~\cite{Biancalani2014} for the model with two states, this noise is strongest in the interior of phase space, and drives the system away from mixed states. The overall amplitude of this noise scales as $N^{-1/2}$; as a consequence consensus states are more likely in small populations, when the noise is strong.

While this general behavior is well-known in the two-species noisy voter model, our analysis focuses on the case of multiple species, and specifically includes models with heterogeneity in the properties of the different species. We find that the transition persists, and we devise a method with which to calculate marginals of the multi-species stationary distribution. Our approach is based on formulating a reduced  stochastic process for single species. This reduction is exact if imitation and mutation rates are homogeneous across species; in the heterogeneous model it is an approximation. 
 We use the effective processes for individual species to obtain analytical results for marginals of the stationary distribution of the model. This allows us to identify the  population sizes $N_L$ and $N_R$, which characterise the transition from a unimodal to a multi-modal stationary distribution. The effective dynamics for single species can also be used to approximate the typical switching times between different consensus states.

There are some limitations to our work. The most severe restriction is that, in its present form, the approach requires all-to-all interaction; this simplifies the analysis greatly. It is an obvious challenge to extend the results we have obtained to multi-state noisy voter models on lattices and more complex graphs.
For example, it would be interesting to see if and how our ideas can be combined with pair-approximation methods used recently for noisy voter models in \cite{Peralta2018}. A further limitation of our approach consists in its inability to capture correlations between species. The method focuses on the marginal dynamics of individual species, and so by construction correlations are not part of the analysis. In order to include these, one next step may be to consider the effective dynamics of pairs of species, while lumping together the remaining species. This would generate an effective three-species model; it is not immediately clear how to proceed with the analytical calculation of joint distributions or switching times. However, progress might be possible in the homogeneous case and exploiting symmetries, similar for example to what has been done in the context of cyclic dynamics in \cite{Cremer2008, Mobilia2010}.

Despite these limitations, we believe our work can contribute to future research in several ways. For example, the method of decoupling multi-species dynamics into several effective single-species processes is likely to be applicable more widely. This includes other models of opinion dynamics and language evolution. In many of these instances evolution is `neutral', i.e., no species has an intrinsic advantage over any other. In such homogeneous models, an approach based on effective single-species processes can be expected to produce accurate results. On the other hand, there are also situations in which heterogeneity is relevant, including evolutionary processes in biology.  The method we have put forward can then be useful as an approximation.


\section*{Acknowledgements}\vspace{-.2cm}
FHA thanks Consejo Nacional de Ciencia y Tecnolog\'ia (CONACyT, Mexico) for support.


\begin{appendix}

\section{Diffusion approximation}\label{sec:diffapprox}
Insight into the model can be gained by approximating the discrete degrees of freedom by continuous variables, and the stochasticity in the dynamics as Gaussian noise. This is based on an expansion in the inverse size of the population, known as the diffusion approximation in mathematical biology \cite{Hartl2007}. In this Appendix we briefly summarise the outcome of this approximation for the multi-state noisy voter model. 

\subsection{Kramers--Moyal expansion and stochastic differential equation}\label{sec:SSExpansion}
Formally, the approximation results in a set of coupled stochastic differential equations (SDEs) for the variables $x_i=n_i/N$. These can be obtained by carrying out a Kramers--Moyal expansion of the master equation~(\ref{eq:master}) in the limit of large, but finite population sizes $N$. Alternatively, these SDEs can also be written down using Kurtz' theorem \cite{Kurtz1978}, or Gillespie's derivation of the chemical Langevin equation for reaction systems \cite{Gillespie2000}. 

For the multi-state noisy VM, as defined by Eqs.~(\ref{eq:master}) and (\ref{eq:rates}), one obtains 
	\begin{equation}\label{eq:sde}
	\dot x_i = f_i(\bx)+\frac{1}{\sqrt{N}}\xi_i(t),
	\end{equation}
where the $\{\xi_i(t)\}$ are zero-average Gaussian noise variables, and where
	\begin{equation}\label{eq:fx}
	f_i(\bx)=\sum_j [{\cal T}_{j\to i}(\bx)-{\cal T}_{i\to j}(\bx)].
	\end{equation}
We have introduced the notation ${\cal T}_{j\to i}(\bx)\equiv T_{j\to i}(N\bx)/N$. The $f_i(\bx)$ are commonly referred to as `deterministic drift' \cite{Risken1996}. They describe the flow of the dynamics in the limit of infinite populations, $N\to\infty$. In this limit intrinsic stochasticity reduces to zero, due to the pre-factor $N^{-1/2}$ in Eq.~(\ref{eq:sde}). The noise variables $\xi_i(t)$ are uncorrelated in time, but correlated across components. Writing
	\begin{equation}
	\avg{\xi_i(t)\xi_j(t')}=B_{ij}(\bx)\delta(t-t'),
	\end{equation}
one has, for example based on Kurtz' theorem \cite{Kurtz1978},
	\begin{equation}\label{eq:bmatrix}
	B_{ij}(\bx)=\left\{\begin{array}{cc}\sum_{k\neq i} \left[{\cal T}_{i\to k}(\bx)+{\cal T}_{k\to i} (\bx)\right]& \mbox{if~}i=j \\
	~\\
	-\left[{\cal T}_{j\to i}(\bx)+{\cal T}_{i\to j}(\bx)\right] & \mbox{if~}i\neq j\end{array}\right. .
	\end{equation}
See also \cite{Bladon2010} for further details in a different context. We note that $\sum_j B_{ij}=0$ for all $i$, reflecting the fact that the total number of individuals in the population is constant.

In explicit form, one finds the drift terms for the multi-state noisy VM as follows,
\begin{equation}
 f_i(\bx)= \sum_{j\neq i}\left[x_i x_j (r_{ij}-r_{ji})+\varepsilon_{ji}x_j - \varepsilon_{ij}x_i\right],\label{eq:Sf_i}
\end{equation}
and the correlation matrix (across components) for the variables $\xi_i(t)$ has entries
	\begin{equation}\label{eq:SB_ij}
	B_{ij}(\bx) =	-x_ix_j\left(r_{ij}+r_{ji}\right)-\varepsilon_{ij}x_i-\varepsilon_{ji}x_j,
	\end{equation}
for $i\neq j$, and 
	\begin{equation}\label{eq:bii}
	B_{ii}(\bx)=-\sum_{j\neq i} B_{ij}(\bx).
	\end{equation}
These results are valid both in the case of homogeneous and heterogeneous imitation and mutation rates across species.

\subsection{Homogeneous rates}
We now focus on individual species in the model with homogeneous rates. The Fokker-Planck equation governing the marginal distribution for species $i$ is given by
	\begin{align}\label{eq:FPEmarg}
	\frac{\partial {\cal P}_i(x_i)}{\partial t} =& - \frac{\partial}{\partial x_i} \left[  f_i(x_i){\cal P}_i(x_i)\right]\nonumber \\
	&+\frac{1}{2N}\frac{\partial^2}{\partial x_i^2} \left[  B_{ii}(x_i){\cal P}_i(x_i)\right],
	\end{align}
where $f_i(x_i)$ and $B_{ii}(x_i)$ are found to depend only on $x_i$ in the homogeneous model. This reduction is similar to the one observed for the transition rates $T_i^\pm$ in Sec.~\ref{sec:hom}. We note that Eq.~(\ref{eq:FPEmarg}) can be obtained from a direct Kramers--Moyal expansion of the master equation describing the death-birth process for species $i$, with rates as in Eq.~(\ref{eq:homrates}).

In explicit form we have
	\begin{equation}
	 f_i(x_i)=\varepsilon(1-m x_i),
	\end{equation}
and 
	\begin{equation}
	B_{ii}(x_i)=\left[\varepsilon(m-2)+2r(1-x_i)\right]x_i+\varepsilon.
	\end{equation}
In the homogeneous model these are the same for all species.
 
The stationary solution of Eq.~(\ref{eq:FPEmarg}) is found as
	\begin{equation}\label{eq:marginal}
	{\cal P}_i(x_i)=C_i\exp\left[\int_0^{x_i}dy~\frac{2 f_{i}(y)-N^{-1} B'_{ii}(y)}{B_{ii}(y)}\right],
	\end{equation}
where $B_{ii}'(y)$ is the derivative of $B_{ii}(y)$ with respect to $y$, and where $C_i$ is a constant ensuring normalisation.
 
We note that the derivative of the marginal stationary distribution with respect to $x_i$ is given by
	\begin{equation}
	{\cal P}_i'(x_i)=C_i\left[2 f_{i}(x_i)-N^{-1} B'_{ii}(x_i)\right]/B_{ii}(x_i).
	\end{equation}
The conditions used to define $N_{L}$ and $N_{R}$ translate into ${\cal P}_i'(0)=0$ and ${\cal P}_i'(1)=0$, respectively. We therefore find
	 \begin{equation}\label{eq:nlfpe}
	N_{L}=\frac{ B'_{ii}(0)}{2 f_{i}(0)}, ~~~ N_{R}=\frac{ B'_{ii}(1)}{2  f_{i}(1)},
	\end{equation}
within the diffusion approximation.

From Eq.~(\ref{eq:nlfpe}) one then obtains
	\begin{align}
	N_L =& \frac{2+(m-2)\varepsilon/r}{2\varepsilon/r}, \nonumber \\
	N_R =& \frac{2-(m-2)\varepsilon/r}{2(m-1)\varepsilon/r}. \label{eq:nlfpe2}
	\end{align}
Results from Eqs.~(\ref{eq:nlfpe2}) are shown as dashed lines in Fig.~\ref{fig:CriticalN}.

In the limit of small mutation rates, $\varepsilon\ll r$, we find
	\begin{equation}
	N_L\approx \frac{r}{\varepsilon}, ~ N_R\approx\frac{r}{(m-1)\varepsilon}.
	\end{equation}
This reproduces the asymptotic behaviour of $N_L$ and $N_R$ in Eqs.~(\ref{eq:nl}) and (\ref{eq:nr}).

The results given in this section are for the case of homogeneous rates; one can then formulate a closed Fokker-Planck equation for the marginal distribution of single species [Eq.~(\ref{eq:FPEmarg})]. If the imitation and mutation rates vary across species, this is no longer possible in exact form. However, an approximate closed Fokker-Planck equation can be formulated for ${\cal P}_i$, following the principles of Sec.~\ref{sec:aapprox}. The approximation consists of replacing $x_j$, $j\neq i$, with the fixed point values $x_j^*$ of the deterministic dynamics for infinite populations. The analog of Eqs.~(\ref{eq:nlfpe}) can then be obtained following the steps described above.

\subsection{Further criterion to characterise the transition}
In this section we briefly discuss a further method to identify the transition between the regimes with unimodal and multi-modal stationary distributions. This is motivated by the procedure used in \cite{Biancalani2014}. We restrict the discussion to models with homogeneous imitation and mutation rates across species.

The solution in Eq.~(\ref{eq:marginal}) can be written in two equivalent forms,
	\begin{equation}\label{eq:ppm}
	{\cal P}(x)=\frac{D\left[A\pm q(x)\right]^{\mp2\gamma\alpha}}{B(x)^{1-\gamma\left(1\pm\alpha\right)}},
	\end{equation}
where we have dropped the index $i$ for simplicity, and where
	\begin{align}
	A  = & \sqrt{\left(\frac{\varepsilon}{r}\frac{m-2}{2}+1\right)^{2}+2\frac{\varepsilon}{r}}, \nn \\
	q(x)  = & 2x-\left(\frac{\varepsilon}{r}\frac{m-2}{2}+1\right), \nn \\
	\alpha  = & \frac{\left(\frac{\varepsilon}{r}+\frac{2}{m}\right)\left(\frac{m-2}{2}\right)}{A}, \nn \\
	\gamma  = & N\frac{m}{2}\frac{\varepsilon}{r}.
	\end{align}
The constant $D$ in Eq.~(\ref{eq:ppm}) ensures normalisation.
We stress that both forms of ${\cal P}$ in Eq.~(\ref{eq:ppm}) describe the same mathematical expression. The purpose of giving both representations will become clear below. We also note that the only dependence of ${\cal P}(x)$ on $N$ is in $\gamma$, and the only dependence
on $x$ is in $q(x)$ and $B(x)$.

In ref.~\cite{Biancalani2014} a similar form of the stationary distribution is obtained for the model with two species; our result reduces to this known case when $m=2$. We note that the dependence on $x$ in the numerator of Eq.~(\ref{eq:ppm}) vanishes for $m=2$, as $\alpha=0$ in this case. In ref.~\cite{Biancalani2014} the phase transition for the two-species model was identified as the population size $N$ for which the stationary distribution is flat; for $m=2$, this is equivalent to setting the exponent of $B(x)$ in Eq.~(\ref{eq:ppm}) to zero, leading to $\gamma=1$, i.e., $N_c=r/\varepsilon$.

We propose a similar criterion to characterise the transition for the multi-state model: we find the population size at which the exponent $1-\gamma(1\pm\alpha)$ in the denominator of Eq.~(\ref{eq:ppm}) vanishes, notwithstanding the fact that a residual dependence on $x$ remains through $q(x)$.

Proceeding on this basis, we use
	\begin{equation}\label{eq:helpme}
	\gamma\left(1\pm\alpha\right)-1 =0
	\end{equation}
to characterise the population size associated with the transition between the unimodal and multi-modal regimes. We note that solving Eq.~(\ref{eq:helpme}) for $N$ leads to two different solutions, one for each choice of the sign in front of $\alpha$. We label these as $N_{\pm}$, and find
	\begin{equation}\label{eq:N+-}
    	N_-=\frac{2r}{m\varepsilon}\frac{1}{1-\alpha},
	~~~ N_+=\frac{2r}{m\varepsilon}\frac{1}{1+\alpha},
	\end{equation}
which for $\varepsilon/r\ll1$ simplifies to
	\begin{equation}\label{eq:N+-approx}
	N_-\approx\frac{r}{\varepsilon}, ~~~ N_+\approx\frac{r}{\varepsilon}\frac{1}{m-1}.
	\end{equation}
To see this note that $A\approx 1$, and $\alpha\approx \frac{m-2}{m}$ in this limit. Eqs.~(\ref{eq:N+-approx}) reproduce the asymptotic behaviour of the expressions for $N_L$ and $N_R$ in Eqs.~(\ref{eq:nlfpe2}) for $\varepsilon/r\ll1$.
Results from Eqs.~(\ref{eq:N+-}) are shown as dotted lines in Fig.~\ref{fig:CriticalN}.

\section{General heterogeneous model}\label{sec:genhet}
We briefly address the model with general heterogeneous imitation and mutation rates, $r_{ij}$ and $\varepsilon_{ij}$. 

Using the transition rates in Eq.~(\ref{eq:rates}), the effective birth and death rates for species $i$ in Eq.~(\ref{eq:tpm}) can be written as
	\begin{align} \label{eq:tpmapp}
	T_{i}^+(\bn)=&\frac{n_i(N-n_i)}{N}\rho_i(\bn) +(N-n_i)\mu_i(\bn),			\nn \\
	T_{i}^-(\bn)=&\frac{n_i(N-n_i)}{N}\lambda_i(\bn) +n_i (m-1)\nu_i,
	\end{align}
where
	\begin{align} \label{eq:averages}
	\rho_i(\bn)		=& \frac{1}{N-n_i}\sum_{j\neq i} n_j r_{ji}, 			\nn \\
	\mu_i(\bn)		=& \frac{1}{N-n_i}\sum_{j\neq i} n_j \varepsilon_{ji}, 	\nn \\
	\lambda_i(\bn)	=& \frac{1}{N-n_i}\sum_{j\neq i} n_j r_{ij}, 			\nn \\
	\nu_i			=& \frac{1}{m-1}\sum_{j\neq i} \varepsilon_{ij}.
	\end{align}

The averages $\rho_i(\bn)$, $\mu_i(\bn)$ and $\lambda_i(\bn)$ are similar to those in Eq.~(\ref{eq:r*-i}). The only complication is a second index of the object which is being averaged. The object $\nu_i$ is a uniform average of $\varepsilon_{ij}$ over all species $j\neq i$, and not dependent on the state $\bn$ of the population.

As in the main text, one can proceed by replacing $n_i/N$ with the values $x_i^*$ at the deterministic fixed point. This approximation results in closed definitions for death-birth processes for individual species $i$. These can then be analysed following the same steps as in Sections~\ref{sec:StatDist}~and~\ref{sec:SwitchTimes}.
\end{appendix}


%

\end{document}